\documentclass[aps,prb,twocolumn,showpacs,groupedaddress,amsmath,amssymb]{revtex4-2}
\usepackage{amsmath}
\usepackage{amssymb}
\usepackage{graphicx}
\usepackage{bbm}
\usepackage{epsfig}
\usepackage[usenames]{color}
\usepackage{bm}
\usepackage{amsthm}
\usepackage{mathrsfs}
\usepackage{float}
\usepackage{subfigure}
\usepackage[section]{placeins}
\usepackage[colorlinks,
linkcolor=blue,
anchorcolor=blue,
citecolor=blue]{hyperref}

\newcommand{\ve} {\varepsilon}

\newcommand{\be}{\begin{eqnarray}}
	\newcommand{\ee}{\end{eqnarray}}
\newcommand{\la}{\langle}
\newcommand{\ra}{\rangle}
\newcommand{\rar}{\rightarrow}
\newcommand{\da}{\downarrow}
\newcommand{\ua}{\uparrow}

\begin{document}
	\title{Unique and Universal scaling  in dynamical quantum phase transitions }
	\author{Xiang Zhang$^1$}
		\author{Liangdong Hu$^2$}
	\author{Fuxiang Li$^1$}
	\email[Corresponding author: ]{fuxiangli@hnu.edu.cn}
	\affiliation{$^1$School of Physics and Electronics, Hunan University, Changsha 410082, China}
	\affiliation{$^2$Department of Physics, School of Science, Westlake University, Hangzhou 310030, China.}
	\date{\today,\now}
\begin{abstract}
	 Universality and scaling are fundamental concepts in equilibrium continuous phase transitions. Here,  we unveil a unique and universal scaling behavior of the critical time in slowly driven dynamical quantum phase transition. Going beyond the analogy with equilibrium phase transition, we find that the critical time exhibits a power-law scaling with quenching rate and  the scaling exponent is fully determined by underlining universality class. We explain this unique scaling behavior based on the adiabatic-impulse scenario in the Kibble-Zurek mechanism. 
  This universal  scaling behavior is verified to be valid  not only in noninteracting single-particle system, but also in many-body interacting system, and not only in Hermitian system, but also in non-Hermitian system. Our study unravels a deep and fundamental relationship between dynamical phase transition and equilibrium phase tranition.
\end{abstract}
\date{\today}
\maketitle

{\it Introduction. --}
Driven by the experimental advances in various quantum-simulator platforms such as ultracold atoms\cite{Langen} and trapped ions\cite{Leibfried}, nonequilirbium quantum dynamics has thrived as a central research field in physics. Intriguing out-of-equilibrium phenomena, such as many-body localization\cite{Abanin,Schreiber,Smith,Choi}, quantum Kibble-Zurek mechanism\cite{Zurek2005, Polkovnikov2005, Sen2008, Barankov2008, Dziarmaga2010, Polkovnikov2011, Nowak2021, Kou2023, Liang2024}, time-crystal\cite{Zaletel} have been theoretically predicted and experimentally verified. 
Among these, the idea of dynamical quantum phase transition (DQPT)\cite{Heyl2018,Lang2018,Bandyopadhyay2021,Vajna2015,Bhattacharya2017A,Vajna2014} has been proposed as a powerful framework to characterize quantum dynamics of  nonequilibrium many-body systems. DQPT has been studied extensively both experimentally\cite{Tian,Jurcevic,Zhang,Fläschner,Guo,Wang,Xu2020} and theoretically\cite{Huang2016,Zache2019,Tian2020,Peotta2021,Corps2023,Sehrawat2021,Hagymási2019,Kosior2024,Jafari2024,Halimeh2017,Valentin2017,Homrighausen2017,Johannes2018,Halimeh2020} and have advanced our understanding of nonequilibrium physics in quantum many-body systems, which may lead to potential applications in quantum computing.

In analogy to the thermal phase transition characterized by nonanalyticities at critical temperature in the thermal free energy, DQPT of a quantum many-body system during the temporal evolution happens when the dynamical analog of the free energy exhibits nonanalyticities at critical evolution times $t_c^*$ \cite{Heyl2013}. In this analogy, Loschmidt amplitude corresponds to the equilibrium partition function of a system.   The evolution time plays the role of control parameter and can be understood as imaginary inverse temperature. 
Renormalization group transformations have shown that the DQPTs are critical points associated with unstable fixed points of equilibrium Ising models, indicating that  DQPT obeys scaling and universality \cite{Trapin2021,Heyl2015,Wu2020,Halimeh2021,HalimeharXiv}.
Moreover, a dynamical analog of  order parameter, coined dynamical topological order parameter\cite{Budich2016}, was proposed to characterize  DQPT. Despite these efforts, the deep connection between DQPT and equilibrium phase transition is still not fully explored. Moreover, it is well established that when a system is slowly driven across a continuous phase transition,  topological defects\cite{Uhlmann2007,Uhlmann2010,Uhlmann2010b,Schützhold2006} will be produced and the density of defects exhibit a power-law scaling behavior with respect to quenching rate \cite{Damski2005}. The scaling exponent is determined by the critical exponents of corresponding equilibrium phase transition. It is natural to ask whether there exists such a quantity in DQPT to connect the dynamics to equilibrium phase transition, and what role the evolution time plays in nonadiabatic dynamics.

In this Letter, we unveil a fundamental relation between DQPT and the equilibrium phase transition. By slowly quenching the system, we find that the critical time $t_c^*$ of DQPT exhibits a universal scaling power law with respect to the quenching rate $\beta$: 
\be
t_c^* \sim \beta^{-\sigma}, ~~{\rm with}~~ \sigma=\frac{\nu z}{1+\nu z}. \label{tc}
\ee
Here, $z$ and $\nu$ are the dynamical exponent and critical exponent of correlation length corresponding to the equilibrium phase transition. Combining analytical analysis and numerical calculations, this universal relation is verified  in different models belonging to different universality classes. These models range from noninteracting to interacting many-body and from Hermitian to non-Hermitian system. Moreover, this relation (\ref{tc}) shares the same form as the transition time separating the impulse and adiabatic regimes in the argument of Kibble-Zurek mechanism, thus reveals a deep connection between Kibble-Zurek mechanism and DQPT. %




{\it Model and result.--}
We start from the transverse-field Ising model (TFIM) for the study of quench dynamics:  
\be
 H&=-\sum_{i, j} J_{i j}\sigma_{i}^{x}\sigma_{j}^{x}-h\sum_{i}\sigma_{i}^{z}, 
\ee 
where $\sigma_{i}$'s ($i=x, y, z$) are the Pauli matrices, $J_{ij}>0$ is the ferromagnetic  interaction in the $x$ directions, and $h$ denotes the  transverse field strength. We will consider two  different cases of $J_{ij}$. One is the short-range TFIC with $J_{ij}=J$ for nearest neighbor sites only and $J_{ij}=0$ otherwise. The second one is the  fully connected TFIC: $J_{ij}=J/N_s$ for all $i$ and $j$, which is also called Lipkin-Meshkov-Glick (LMG) model. Here, $N_s$ is the total number of spins.
For both cases, the system is in the ferromagnetic phase for  $|h|<J$  and in the paramagnetic phase for $|h|>J$. However, near the critical point $|h_c|=J$, the two models belong to distinctive universality classes. The short-range TFIC has critical exponents $z=\nu=1$, while the LMG model has critical exponents $z=1/3$ and $\nu=3/2$\cite{Botet1983}. Here, $z$ and $\nu$ are the dynamical exponent and critical exponent of correlation length. We will show that under slow quench dynamics, the critical time of DQPT exhibits different scaling exponents for the two different universality classes. In Supplemental Material (SM) \cite{sm}, other cases of $J_{ij}$ are also discussed.

\begin{figure}
    \includegraphics[width=1\linewidth]{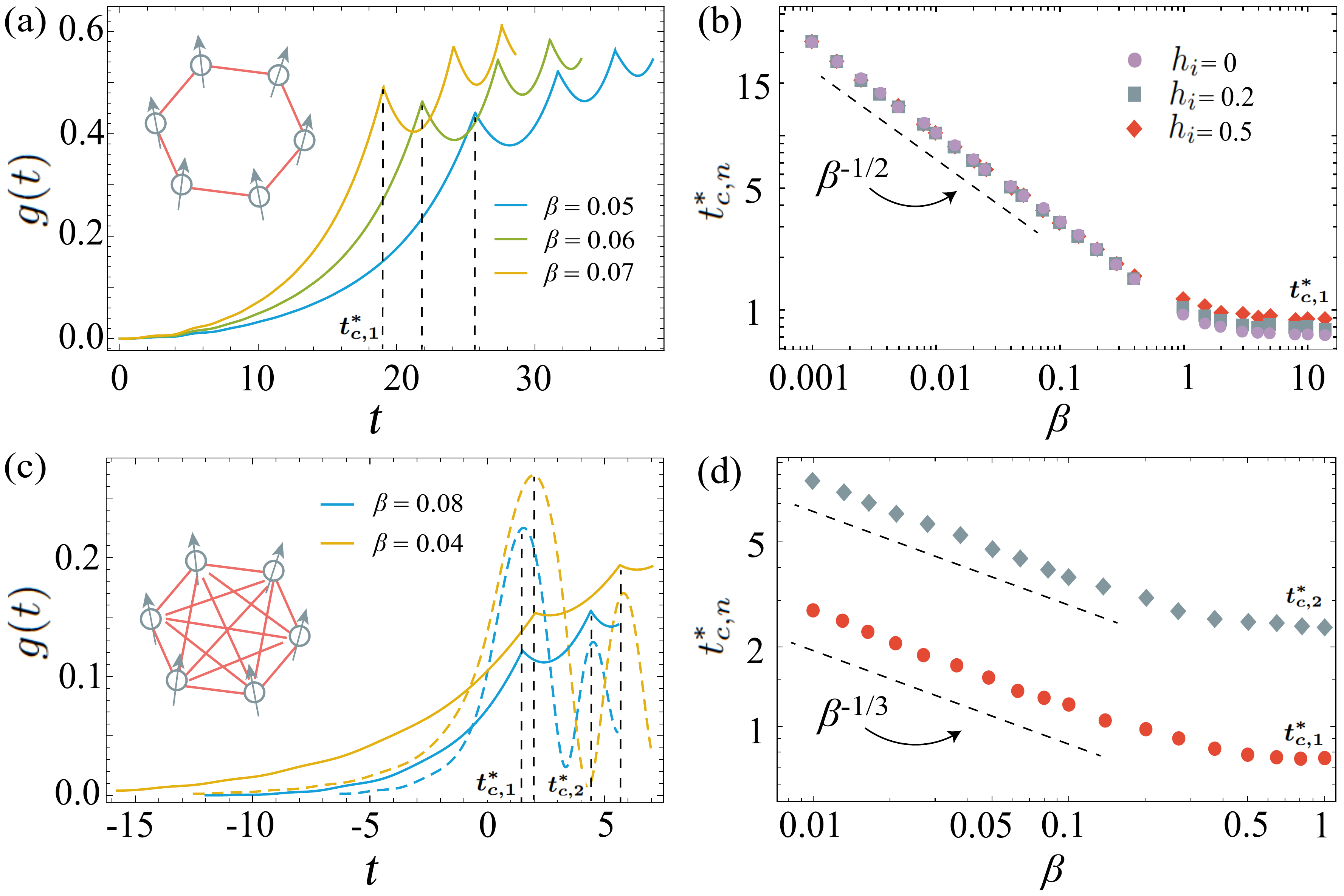}
	\caption{ DQPT and scaling laws in the short range TFIC (a-b) and LMG model (c-d) under slow quench dynamics.  (a) and (c) Rate function $g(t)$ for different quenching rates $\beta$.  The solid lines correspond to exact numerical calculations and the dashed lines correspond to elementary excitation method. The vertical dotted lines correspond to the first critical times $t^*_{c,1}$.  (b) The first critical time as a function of quenching rate $\beta$ for different initial values $h_i$.  (d) The first and second critical time  vs. quenching rate $\beta$ for LMG model obtained using exact numerical calculation. }
	\label{TFIC}
\end{figure}

We take the slow quench protocol as $h(t)=1+\beta t $ and $J=1$ for the short-range TFIM, and $J= 1-\beta t$  and $h=1$ for the  LMG model, respectively, with time varying from initial  $t_{i}=-1/\beta$ to final  $t_{f}=1/\beta$ \cite{Puškarov2016,Sharma2016,Zamani,Cao,Baghran}. 
 The  quench protocols correspond to  a dynamical transition from ferromagnetic to paramagnetic phases.  
Starting from the ground state $|\Psi(t_i)\ra$ at initial time $t_i$, the wave function can be obtained by solving the Schr${\rm \ddot{o}}$dinger equation $i\partial_t \Psi(t)=H(t)\Psi(t) $ governed by the time-dependent Hamiltonian $H(t)$. 
The key physical quantities in the theory of DQPT are the Loschmidt amplitude (LA)
$G\left(t\right)=\langle \Psi(t_i)|\Psi(t)\rangle$ and the rate function 	
\be
g\left(t\right)=-\lim_{N\rightarrow\infty}\frac{1}{N}\ln |G\left(t\right)|^{2},
\ee 
where $N$ is the number of degrees of freedom. In analogy to equilibrium phase transitions where they are associated with singular behavior in free energy, DQPT appears when $g\left(t\right)$ becomes nonanalytic at critical times.

As shown in Fig.~\ref{TFIC}, for both models, rate function $g(t)$ exhibits singular points at some critical times. To study the universality and scaling of DQPT, we focus on the critical time $t^*_{c,n}$ rather than the value of rate function. We discover that the critical time $t^*_{c,n}$ exhibits a power-law scaling behavior with respect to quenching rate $\beta$ for sufficiently slow quench dynamics. However, for the two models, the scaling exponents $\sigma$ are different. For short-range TFIC, $\sigma =1/2$,  while for LMG model,  $\sigma=1/3$.  Below we will argue that the scaling exponent is deeply connected to the universality class of the critical phase transition, and is fully determined by critical exponents of equilibrium phase transition.

{\it Short Range TFIM.--}We first study DQPT in the short-range TFIC. Under periodic boundary condition, we employ Jordan-Wigner transformation to map spins to spinless fermions and apply Fourier transformation to the momentum space in which the Hamiltonian are decoupled for each mode $k$ with each sector described by the following $2\times 2$ Hamiltonian \cite{sm}
\be
 		H_{k} = (-h+\cos k) \sigma_z + \sin k \sigma_x.
\ee
 Diagonalizing the Hamilitonian, we get the eigenvalues  $E_k=\pm\varepsilon_k=\pm\sqrt{(h-\cos k)^{2}+\sin^2 k}$.
When $h = \pm 1$,  the gap vanishes for momentum values $k = 0$ and  $k = \pi $, respectively. 
Because the Hamiltonian is decoupled in $k$ space, the LA can be written as 
\be
		G(t)=\prod_k G_k(t)=\prod_k\la \psi_k(0)|\psi_k(t)\ra.
\ee 
Here, wave function $|\psi_k(t) \ra$ for each $k$ is obtained by solving the time-dependent Schr${\rm \ddot{o}}$dinger equation under time-dependent Hamiltonain $H_k(t)$. 

One notices that, since the energy gap closes at $k=0$ when $h=1$, the dynamical process is dominated by the modes near $k=0$. Therefore, one can safely approximate the Hamiltonian $H_k(t)$ by keeping only terms of the leading order of $k$. With a proper rotation of spin indices, the problem reduces to a standard Landau-Zener (LZ) problem: $i\partial_t \psi_k(t) = (-\beta t \sigma_z + k \sigma_x) \psi_k(t) $ with the spinor $|\psi_k(t)\ra=(a(t), b(t))^T$. The initial condition can be set to $|\psi_k(t_i)\ra = (1, 0)^T$ for $t_i \rar \infty$.


To understand the scaling behavior, we first review the solution of LZ model. For this standard LZ problem, one first  obtains a seconder order differential equation of $a(t)$, which can be reduced to the standard parabolic cylinder equation if one adopts two dimensionless parameters \cite{sm}
\be
&&\alpha =   k/\sqrt{2\beta}, ~~
\xi = \sqrt{\beta/2} t. \label{eqvariable}
\ee
The solution $|\psi_k(t)\ra$ is a combination of  two linearly independent parabolic functions $D_{i\alpha }(z)$ and $D_{i\alpha}(-z)$ with $z=2 \xi e^{-i\pi/4}$. Here, simply from the dimensionless parameter (\ref{eqvariable}), it is easy to see that the critical time should scale with $\beta$ in such a way $t_c^* \sim \beta^{-1/2}$, since the DQPT happens at some fixed parameter $\xi^*$.

Now, instead of exactly solving the LZ problem, we resort to a simple argument to obtain a more physical picture of the scaling behavior.  Based on Kibble-Zurek mechanism, the LZ transition can be considered to consist of  two regimes, the adiabatic and impulse regime \cite{Damski2005}. At initial time (large negative time), the system can adiabatically follow the ground state $|\da (t)\ra$ until at the transition time $\hat{t}$ when the relative change rate of control parameter is equal to  the relaxation time of the system $\tau(t)=1/\ve_k(t) $. Thereafter, the state is effectively frozen to the state at the transition time $|\da(-\hat{t} ) \ra$. The transition time can be estimated to be $\tau=c_0 \hat{t}$ with  $c_0$ being a fitting parameter of the order of unity. 
The frozen regime ends at time $\hat{t}$, and the system then enters into the adiabatic region, in which the  probability remains unchanged on the upper and lower adiabatic eigenstate and only the phase is accumulated. Therefore, at time $t>\hat{t}$, one can write down the wave function as:
\be
|\psi(t)\ra = \sqrt{P_{\ua}} e^{-i \eta (t) } |\ua(t)\ra + \sqrt{1-P_{\ua}} e^{i \eta (t) } |\da(t)\ra \label{eqvector}
\ee
Here, $P_{\ua}=e^{-\pi k^2/\beta}$  is the Landau-Zener transition probability from initial ground state to the upper  state. $|\ua, \da(t)\ra $ are the instantaneous eigenstates of the time-dependent Hamiltonian.  
$\eta(t)$ is the dynamical phase accumulated on the upper instantaneous state: 
\be
&&\eta(t) = \int^t dt \ve_k(t) = 2 \int^{\xi} d\xi \sqrt{\xi^2 + \alpha^2}  \label{eta}
\ee
where we have made a variable change according to (\ref{eqvariable}).

Instead of making a direct integration of the above integral, we want to argue that the integration can be well approximated by disregarding the $\alpha$ term and obtain  $\eta (t) \sim \xi^2$. According to (\ref{eqvector}), the first DQPT occurs  when two conditions are satisfied: $P_{\ua}=1-P_{\ua}$ and $\eta(t^*_{c,1}) = \pi/2$. The first condition  leads to  $\alpha=\sqrt{k^2/(2\beta)}=\sqrt{\ln2/2\pi} \approx 0.33$. The second condition leads to $\xi_c^* \sim \sqrt{\pi/2}\approx 1.14$ and thus the scaling relation $t^*_{c,1} \sim \sqrt{\pi}/\sqrt{\beta}$.  The approximation is thus guaranteed by the fact that $\xi_c^* \gg \alpha$.  Two remarks are in order. 
First, we always have $\xi_c^*> \hat{\xi}$. Therefore, DQPT occurs after the frozen time $\hat{t}$, and the adiabatic approximation (\ref{eqvector}) is valid.  
Second, the first condition  also leads to a scaling of $k^*$ that dominates the DQPT  $k^* \sim \sqrt{\beta}$, which has been discussed in Ref.\cite{Puškarov2016,Zamani}.  

 The above argument leads to a simple but general explanation of the scaling behavior that is also applicable to many-body system. 
For a many-body system, the energy gap between ground state and first excited state (or the most relevant state)  is
  \be
 \Delta(t) \sim (\beta t)^{\nu z}.\ee
According to the adiabatic-impulse picture,   the instantaneous many-body ground state $|\Psi_0\ra$ and the excited state $|\Psi_1(t) \ra$ should play the dominating roles in the dynamical processes, and should replace the two-level states $|\ua, \da\ra$ in the total wavefunction $|\psi(t) \ra$ in (\ref{eqvector}). Now the relative phase factor between the ground state and excited state is
 \be
 \eta(t)  = \int dt \Delta(t) = \frac{1}{1+\nu z} \beta^{\nu z} t^{\nu z+1}
 \ee
The DQPT accurs at $t^*_{c, n}$ such that $\eta(t^*_{c, n}) = (2n+1) \pi/2$ with $n=0, 1, 2, \ldots$. We have the following scaling relation
 \be
t_{c,n}^* =c_n \beta^{-\frac{\nu z}{1+ \nu z}}
 \ee
 with $c_n= [(1+\nu z)(n+1/2)\pi]^{1/(1+\nu z)} $. 
 
We  note that the critical time relative to the time approaching the static critical phase point (time $t=0$), is not affected by $h_i$ and $h_f$ for slow quench dynamics as shown in Fig.~\ref{TFIC}(b). This is because the dominant process happens near the transition time from adiabatic regime to impulse regime, while $h_i$ and $h_f$ are far away from this transition time.

{\it LMG model.--} The time-dependent Hamiltonian of the fully connected LMG model\cite{Dusuel2005,Acevedo2014,Titum2020}  can be rewritten in terms of a single collective spin of length $N/2$, $S_x=\sum_{i}\sigma_i^x/2$, $S_z=\sum_{i}\sigma_i^z/2$ such that
\begin{equation}
	H=-\frac{J(t)}{N}S_x^2-S_z \label{eqHS}
\end{equation}
Adopting the quench protocol $J(t)=1-\beta t$, the system is set to be in the ground state $|\rar\ra$ at initial time $t_i=-1/\beta$. Here, $|\rar\ra $ is the eigenstate of $S_x$ with the maximum eigenvalue. The scaling behavor of critical time can be well explained if one adopts the critical exponents $\nu=3/2$ and $z=1/3$. 

One can understand the scaling behavior of DQPT by considering the elementary excitations\cite{Dabrowski2016,Lewis1967,Defenu2018}. By applying a Holstein-Primakoff transformation followed by a Bogoliubov transformation, we can approximate the Hamiltonian (\ref{eqHS}) in terms of  bosonic operators $b$ and $b^\dagger$:
\begin{equation}
	\label{lmg}
	H=\Omega(t) b^\dagger b.
\end{equation}
The Hamiltonian (\ref{lmg}) corresponds to a single harmonic oscillator with time-dependent frequency $\Omega(t)$. For the linearly quenched Hamiltonian  $\Omega^2(t)= \beta t$ in the large $N$ limit. The time evolution of each harmonic oscillator can be exactly solved and is fully described by a single complex parameter, the effective width $\xi(t)$, which satisfying the Ermakov-Milne equation 
\be
\ddot{\xi}(t)+\Omega^2(t)\xi(t)=\frac{1}{4\xi^3(t)}.  \label{xi}
\ee
The rate function $g(t)$ obtained from the wavefunction of  harmonic oscillator  is labelled as dotted lines, shown in Fig.~\ref{TFIC}(c). The analytical results of LA is shown in SM \cite{sm}. It can be seen that, no singularity appears. However, it still exhibits several peaks that are at the same critical times $t^*_{c,n}$ with the DQPT obtained from exact numerical calculations. To understand the scaling behavior, one can resort to Eq.~(\ref{xi}). To solve (\ref{xi}), one first needs to make it dimensionless by making the following change of parameters: 
\be
\xi = \beta^{-1/6} \tilde{\xi}, ~~ t=\beta^{-1/3} s. 
\ee 
Again, as in the case of short-range TFIM, DQPT occurs at some fixed value of dimensionless parameter $s$, which gives rise to the scaling of $t^*_c \sim \beta^{-1/3}$. 



\begin{figure}[htbp]
	\includegraphics[width=1\linewidth]{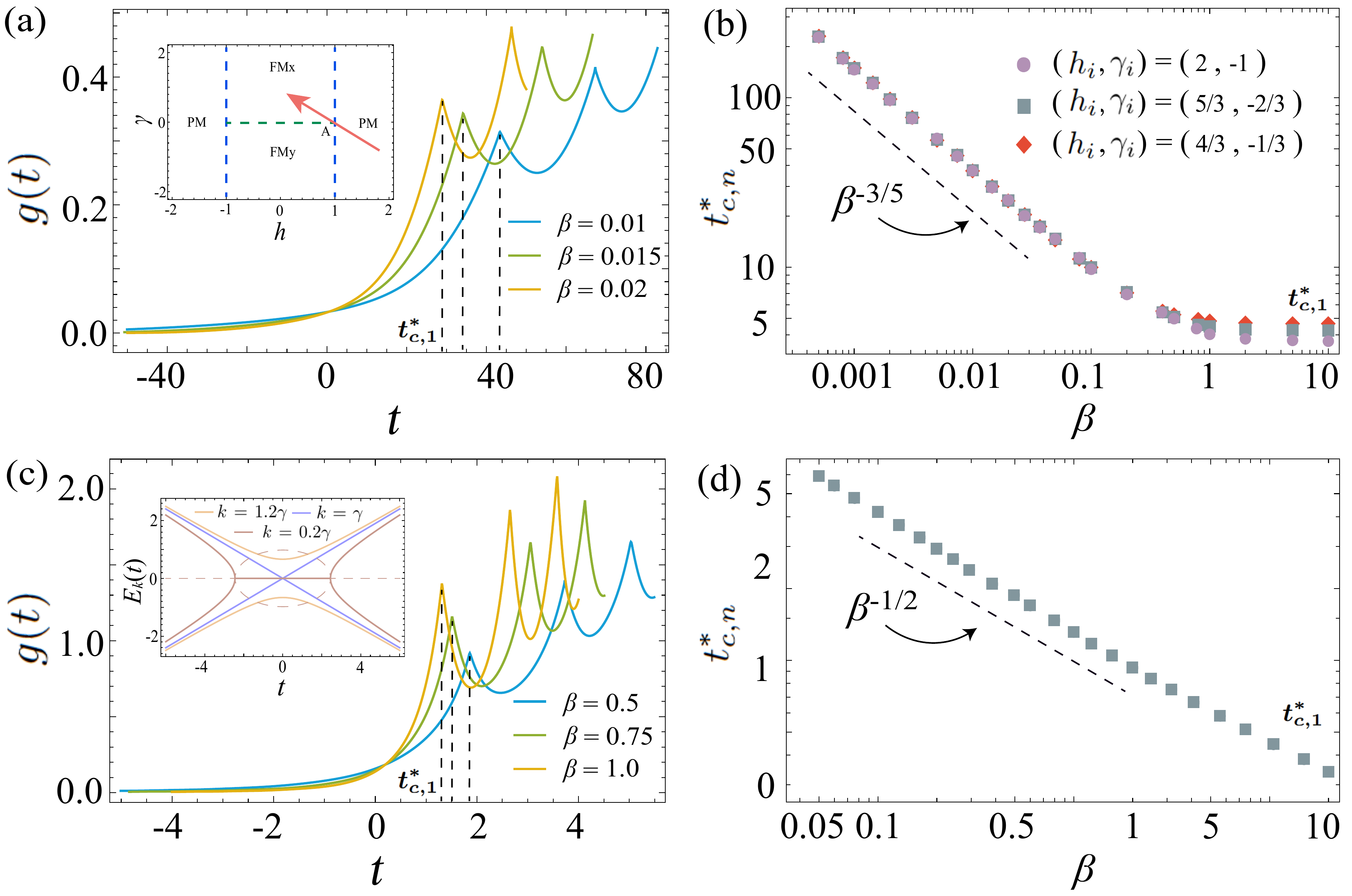}
	\centering
	\caption{Rate function $g(t)$ of the Loschmidt amplitude and scaling laws in XY chain (a-b) and non-Hermitian model (c-d). For XY chain, the quench protocol is denoted by the red arrow in the phase diagram.   Inset of (c) plots the instantaneous energy for different values of $k$. Solid and dashed lines denote the real and imaginary parts of energy, respectively. The first critical time scales with quenching rate with exponent $3/5$ and $1/2$, respectively, for XY chain and non-Hermitian model. } 
	\label{nh}	
\end{figure}

{\it XY chain.--} To further verify our findings of the universal scaling of critical time in DQPT, we provide two more examples. The first one is the spin-1/2 anisotropic XY model \cite{Mukherjee2010,Deng2009,Divakaran2008,Porta2020}
\begin{equation}
	H=-\frac{1}{2}\sum_{j=1}^{N}\left(J_{x} \sigma_{j}^{x} \sigma_{j+1}^{x}+J_{y} \sigma_{j}^{y} \sigma_{j+1}^{y}+h \sigma_{j}^{z}\right)
\end{equation}
where  $J_{x}$ and $J_y$ are the anisotropy interactions along $x$ and $y$ directions. A more general model is studied in SM\cite{sm}. For simplicity, we set $J=J_x+J_y$ and $\gamma=(J_x-J_y)/J$. Applying the Jordan-Wigner transformation to map  the spin operators to spinless fermionic operators and making  the Fourier transformation to momentum space, the Hamiltonian can be diagonalized into $2$ by $2$ sectors. 
The energy spectrum of the system is $E_k=\pm \sqrt{(h+J\cos k)^2+J^2\gamma^2\sin^2 k}$. The phase diagram of this model is shown in Fig.~\ref{nh}(a) inset.
Here, we consider a  quench path as indicated by the red line arrow shown in inset.  The quench path traverses the multicritical point $A$, and near the critical point, the critical exponents are $z=2$ and $\nu=1/2$ \cite{Deng2009}.   
In Fig.~\ref{nh}(a), one plots the rate function $g(t)$ for different values of quenching rate $\beta$, and the first DQPT occurs at critical time $t^*_{c, 1}$. In Fig.~\ref{nh}(b), we show that the critical time $t^*_{c,1}$ exhibits a scaling behavior with quenching rate $\beta$ with exponent $3/5$. It is interesting to note that this exponent  cannot be explained by Eq.~(\ref{tc}) together with the static exponent $z=2$. Rather, one has to adopt the effective dynamical critical exponent $z_2=3$ which is determined by the scaling of a path-dependent minimum gap, and was introduced in Ref.~\cite{Deng2009}  to explain the failure of Kibble-Zurek scaling in a multicritical quantum quench. Here, the necessity of introducing the effective exponent $z_2$ reflects the deep connection between slow-quench DQPT and Kibble-Zurek mechanism. 



{\it Non-Hermitian System.--}
DQPT also appears in the non-hermitian system under sudden quench protocol\cite{Zhou2018,Jing2024}. Here, we consider an exactly solvable model of effective two-level systems parameterized by momentum $k$. This is given by the Hamiltonian\cite{Shen2019,Sim2023}
\begin{equation}
	H_k=k\sigma_x+i\gamma \sigma_y+\lambda\sigma_z
\end{equation}
The quench protocol is $\lambda =\beta t$ from $-\infty$ to $+\infty$. This Hamiltonian is a generalization presented in \cite{Lee2015} and realized experimentally
in \cite{Gerritsma2010}, by adding a real drive term $\lambda$ and applying a basis rotation. There, $\gamma$ corresponds to the imaginary tachyon mass, $k$ is the momentum, and $\beta$ is the quenching rate. This Hamiltonian has $\mathcal{PT}$ symmetry,  $\mathcal{P}=\sigma_y$ and $\mathcal{T}=-i\sigma_y\mathcal{K}$ where $\mathcal{K}$ is complex conjugation. At the exceptional point (EP) $k=\gamma$, the spontaneous breaking of this symmetry occurs and the states are no longer eigenstates of the $\mathcal{PT}$ operator. The instantaneous eigenvalues is $\Delta_k(t)=\pm\sqrt{\beta^2t^2+k^2-\gamma^2}$  shown in inset of Fig.~\ref{nh}(c). Near the EP, the Hamiltonian has critical exponent $z\nu=1$.
One can solve the time-dependent Schrodinger equation and obtain the wavefunction exactly (see SM\cite{sm}). As shown in Fig.~\ref{nh}(c-d),  DQPT occurs and the critical time exhibits a power-law scaling with quenching  rate with the exponent  $1/2$ in agreement with the prediction (\ref{tc}).


{\it Discussion and conclusion.--} 
There have emerged various experimental platforms for quantum many-body simulations that may be used to testify our theoretical prediction of universal scaling  within the current technology \cite{Fauseweh2024}.
In the programmable superconducting quantum annealer, the short-range transverse Ising model of up to $1000$ spins has been realized, and a coherent quantum annealing of up to $40$ns can be achieved \cite{King2022}.  Long-range transverse Ising model can be simulated in trapped ion platform and Rydber atom platform and the coherent time in both platforms is sufficiently long so that the quantum Kibble-Zurek mechanism can be testified, which should also enable the test of our prediction under slow quench dynamics \cite{Keesling2019, Xu, Li2023}. 


In summary, our work studies DQPT under slow quench dynamics and discovers a unique scaling behavior of critical time with respect to the quenching rate. The scaling exponent is fully determined by the corresponding equilibrium phase transition. This study thus builds up a one-to-one correspondence between the slow-quench DQPT and the equilibrium phase transition. It is interesting that the underlining physics of the scaling behavior is rooted in the impulse-adiabatic scenario on which the famous Kibble-Zurek mechanism is based, and our results thus also unveil the fundamental relation between DQPT and Kibble-Zurek mechansim.  Moreover, the unique scaling relation of critical time  indicates that time is not simply an analog of imaginary time as in equilibrium phase transition, but  plays a more important role in DQPT under ramped dynamics. 


{\it Acknowledgements.--}
This work was supported by the National Key Research and Development Program of Ministry of Science and Technology (No. 2021YFA1200700), National Natural Science Foundation of China (No. 11905054, and No.11804122), the China Postdoctoral Science Foundation (Grant No. 2021M690970) and the Fundamental Research Funds for the Central Universities from China.


\begin{thebibliography}{100}
    \bibitem{Langen}T. Langen, R. Geiger, and J. Schmiedmayer, Ultracold Atoms Out of Equilibrium, Annu. Rev. Condens. Matter Phys. {\bf6}, 201 (2015).
    \bibitem{Leibfried}D. Leibfried, R. Blatt, C. Monroe, and D.Wineland, Quantum dynamics of single trapped ions, Rev. Mod. Phys. {\bf75}, 281 (2003).

    
	\bibitem{Abanin} D. A. Abanin, E. Altman, I. Bloch, and M. Serbyn, Colloquium: Many-body localization, thermalization, and entanglement, Rev. Mod. Phys. 91, 021001 (2019).
    \bibitem{Schreiber} M. Schreiber, S. S. Hodgman, P. Bordia, H. P. Lüschen, M. H. Fischer, R. Vosk, E. Altman, U. Schneider, and I. Bloch, Observation of many-body localization of interacting fermions in a quasirandom optical lattice, Science {\bf349}, 842 (2015).
    \bibitem{Smith} J. Smith, A. Lee, P. Richerme, B. Neyenhuis, P. W. Hess, P. Hauke, M. Heyl, D. A. Huse, and C. Monroe, Many-body localization in a quantum simulator with programmable random disorder, Nat. Phys. {\bf12}, 907 (2016).
    \bibitem{Choi} J.-y. Choi, S. Hild, J. Zeiher, P. Schauß, A. Rubio-Abadal, T. Yefsah, V. Khemani, D. A. Huse, I. Bloch, and C. Gross, Exploring the many-body localization transition in two dimensions, Science {\bf352}, 1547 (2016).


  
    \bibitem{Zurek2005} W. H. Zurek, U. Dorner, and P. Zoller, Dynamics of a Quantum Phase Transition, Phys. Rev. Lett. {\bf95}, 105701 (2005).
    \bibitem{Polkovnikov2005} A. Polkovnikov, Universal Adiabatic Dynamics in the Vicinity of a Quantum Critical Point, Phys. Rev. B  {\bf72}, 161201 (2005).
    \bibitem{Sen2008} D. Sen, K. Sengupta, and S. Mondal, Defect Production in Nonlinear Quench across a Quantum Critical Point, Phys. Rev. Lett. {\bf101}, 016806 (2008).
    \bibitem{Barankov2008} R. Barankov and A. Polkovnikov, Optimal Nonlinear Passage through a Quantum Critical Point, Phys. Rev. Lett. {\bf101}, 076801 (2008).
    \bibitem{Dziarmaga2010} J. Dziarmaga, Dynamics of a Quantum Phase Transition and Relaxation to a Steady State, Adv. Phys. {\bf59}, 1063 (2010).
    \bibitem{Polkovnikov2011} A. Polkovnikov, K. Sengupta, A. Silva, and M. Vengalattore, Colloquium: Nonequilibrium Dynamics of Closed Interacting Quantum Systems, Rev. Mod. Phys. {\bf83}, 863 (2011).
    \bibitem{Nowak2021} R. J. Nowak and J. Dziarmaga, Quantum Kibble-Zurek Mechanism: Kink Correlations after a Quench in the Quantum Ising Chain, Phys. Rev. B{\bf104}, 075448 (2021).
    \bibitem{Kou2023} H. C. Kou and P. Li, Varying Quench Dynamics in the Transverse Ising Chain: The Kibble-Zurek, Saturated, and Presaturated Regimes, Phys. Rev. B {\bf108}, 214307 (2023).
    \bibitem{Liang2024} En-Wen Liang, Ling-Zhi Tang, and Dan-Wei Zhang, Quantum criticality and Kibble-Zurek scaling in the Aubry-André-Stark model, arXiv:2405.10199 (2024).
   
    
    

    \bibitem{Zaletel} M. P. Zaletel, M. Lukin, C. Monroe, C. Nayak, F. Wilczek, and N. Y. Yao, Colloquium: Quantum and classical discrete time crystals, Rev. Mod. Phys. {\bf95}, 031001 (2023).
    
    
    
    
    \bibitem{Heyl2018} B. Žunkovič, M. Heyl, M. Knap, and A. Silva, Dynamical Quantum Phase Transitions in Spin Chains with Long-Range Interactions: Merging Different Concepts of Nonequilibrium Criticality, Phys. Rev. Lett. {\bf120}, 130601 (2018).
    \bibitem{Lang2018} J. Lang, B. Frank, and J. C. Halimeh, Dynamical quantum
    phase transitions: A geometric picture, Phys. Rev. Lett. {\bf121},
    130603 (2018).
    \bibitem{Bandyopadhyay2021} S. Bandyopadhyay, A. Polkovnikov, and A. Dutta, Observing Dynamical Quantum Phase Transitions through Quasilocal String Operators, Phys. Rev. Lett. {\bf126}, 200602 (2021).
    \bibitem{Vajna2015} S. Vajna and B. Dóra, Topological Classification of Dynamical Phase Transitions, Phys. Rev. B {\bf91}, 155127 (2015).
    \bibitem{Bhattacharya2017A} U. Bhattacharya and A. Dutta, Interconnections between Equilibrium Topology and Dynamical Quantum Phase Transitions in a Linearly Ramped Haldane Model, Phys. Rev. B {\bf95}, 184307 (2017).
    \bibitem{Vajna2014} S. Vajna and B. Dóra, Disentangling Dynamical Phase Transitions from Equilibrium Phase Transitions, Phys. Rev. B {\bf89}, 161105 (2014).

 
 
 
	\bibitem{Tian}T. Tian, Y. Ke, L. Zhang, S. Lin, Z. Shi, P. Huang, C. Lee, and J. Du, Observation of dynamical phase transitions in a topological nanomechanical system, Phys.Rev.B {\bf100}, 024310 (2019).
	\bibitem{Jurcevic} P. Jurcevic, H. Shen, P. Hauke, C. Maier, T. Brydges, C. Hempel, B. P. Lanyon, M. Heyl, R. Blatt, and C. F. Roos,Direct Observation of Dynamical Quantum Phase Transitions in an Interacting Many-Body System, Phys. Rev. Lett. {\bf119}, 080501 (2017).
	\bibitem{Zhang} J. Zhang, G. Pagano, P. W. Hess, A. Kyprianidis, P. Becker, H. Kaplan, A. V. Gorshkov, Z.-X. Gong, and C. Monroe, Observation of a many-body dynamical phase transition with a 53-qubit quantum simulator, Nature (London) {\bf551}, 601 (2017).
	\bibitem{Fläschner} N. Fläschner, D. Vogel, M. Tarnowski, B. S. Rem, D.-S. Lühmann, M. Heyl, J. C. Budich, L. Mathey, K. Sengstock, and C. Weitenberg, Observation of dynamical vortices after quenches in a system with topology, Nature Phys {\bf14}, 265 (2018).
	\bibitem{Guo} X.-Y. Guo, C. Yang, Y. Zeng, Y. Peng, H.-K. Li, H. Deng, Y.-R.Jin,S.Chen, D. Zheng, andH.Fan, Observation of a Dynamical Quantum Phase Transition by a Superconducting Qubit Simulation, Phys. Rev. Appl. {\bf11}, 044080 (2019).
	\bibitem{Wang} K. Wang, X. Qiu, L. Xiao, X. Zhan, Z. Bian, W. Yi, and P. Xue, Simulating Dynamic Quantum Phase Transitions in Photonic Quantum Walks, Phys. Rev. Lett. {\bf122}, 020501 (2019).
	\bibitem{Xu2020} X. Y. Xu et al., Measuring a Dynamical Topological Order Parameter in Quantum Walks, Light Sci. Appl. {\bf9}, (2020).

	

    
	

    \bibitem{Huang2016} Z. Huang and A. V. Balatsky, Dynamical Quantum Phase Transitions: Role of Topological Nodes in Wave Function Overlaps, Phys. Rev. Lett. {\bf}, 086802 (2016).
    \bibitem{Zache2019} T. V. Zache, N. Mueller, J. T. Schneider, F. Jendrzejewski, J. Berges, and P. Hauke, Dynamical Topological Transitions in the Massive Schwinger Model with a $\theta$ Term, Phys. Rev. Lett. {\bf122}, 50403 (2019).
    \bibitem{Tian2020} T. Tian, H. X. Yang, L. Y. Qiu, H. Y. Liang, Y. B. Yang, Y. Xu, and L. M. Duan, Observation of Dynamical Quantum Phase Transitions with Correspondence in an Excited State Phase Diagram, Phys. Rev. Lett. {\bf124}, 043001 (2020).
    \bibitem{Peotta2021} S. Peotta, F. Brange, A. Deger, T. Ojanen, and C. Flindt, Determination of Dynamical Quantum Phase Transitions in Strongly Correlated Many-Body Systems Using Loschmidt Cumulants, Phys. Rev. X {\bf11}, 041018 (2021).
    \bibitem{Corps2023} Á. L. Corps and A. Relaño, Theory of Dynamical Phase Transitions in Quantum Systems with Symmetry-Breaking Eigenstates, Phys. Rev. Lett. {\bf130}, 100402 (2023).
    \bibitem{Sehrawat2021} A. Sehrawat, C. Srivastava, and U. Sen, Dynamical Phase Transitions in the Fully Connected Quantum Ising Model: Time Period and Critical Time, Phys. Rev. B {\bf104}, 085105 (2021).
    \bibitem{Hagymási2019} I. Hagymási, C. Hubig, Ö. Legeza, and U. Schollwöck, Dynamical Topological Quantum Phase Transitions in Nonintegrable Models, Phys. Rev. Lett {\bf122}, 250601 (2019).
    \bibitem{Kosior2024} A. Kosior and M. Heyl, Vortex Loop Dynamics and Dynamical Quantum Phase Transitions in Three-Dimensional Fermion Matter, Phys. Rev. B {\bf109}, L140303 (2024).
    \bibitem{Jafari2024} R. Jafari, A. Langari, S. Eggert, and H. Johannesson, Dynamical Quantum Phase Transitions Following a Noisy Quench, Phys. Rev. B {\bf109}, L180303 (2024).
    \bibitem{Halimeh2017} Jad C. Halimeh and Valentin Zauner-Stauber, Dynamical phase diagram of spin chains with long-range interactions
    Phys. Rev. B {\bf96}, 134427 (2017).
    \bibitem{Valentin2017} Valentin Zauner-Stauber and Jad C. Halimeh, Probing the anomalous dynamical phase in long-range quantum spin chains through Fisher-zero lines, Phys. Rev. E {\bf96}, 062118 (2017).
    \bibitem{Homrighausen2017} Ingo Homrighausen, Nils O. Abeling, Valentin Zauner-Stauber, and Jad C. Halimeh, Anomalous dynamical phase in quantum spin chains with long-range interactions, Phys. Rev. B {\bf96}, 104436 (2017).
    \bibitem{Johannes2018} Johannes Lang, Bernhard Frank, and Jad C. Halimeh,
    Concurrence of dynamical phase transitions at finite temperature in the fully connected transverse-field Ising model, Phys. Rev. B {\bf97}, 174401 (2018).
    \bibitem{Halimeh2020} Jad C. Halimeh, Maarten Van Damme, Valentin Zauner-Stauber, and Laurens Vanderstraeten, Quasiparticle origin of dynamical quantum phase transitions, Phys. Rev. Res. {\bf2}, 033111 (2020).
    
    
    
    \bibitem{Heyl2013} M. Heyl, A. Polkovnikov, and S. Kehrein, Dynamical Quantum Phase Transitions in the Transverse-Field Ising Model, Phys. Rev. Lett. {\bf110}, 135704 (2013).
    
    
    
    \bibitem{Trapin2021} D. Trapin, J. C. Halimeh, and M. Heyl, Unconventional Critical Exponents at Dynamical Quantum Phase Transitions in a Random Ising Chain, Phys. Rev. B {\bf104}, 115159 (2021).
    \bibitem{Heyl2015} M. Heyl, Scaling and Universality at Dynamical Quantum Phase Transitions, Phys. Rev. Lett. {\bf115}, 140602 (2015).
    \bibitem{Wu2020} Y. Wu, Dynamical Quantum Phase Transitions of Quantum Spin Chains with a Loschmidt-Rate Critical Exponent Equal to 1/2, Phys. Rev. B {\bf101}, 064427 (2020).
    \bibitem{Halimeh2021} J. C. Halimeh, D. Trapin, M. Van Damme, and M. Heyl, Local Measures of Dynamical Quantum Phase Transitions, Phys. Rev. B {\bf104}, 075130 (2021).
   \bibitem{HalimeharXiv} J. C. Halimeh, Nikolay Yegovtsev, and Victor Gurarie, Dynamical quantum phase transitions in many-body localized systems,
    arXiv:1903.03109.
    
    
    
    \bibitem{Budich2016} J. C. Budich and M. Heyl, Dynamical Topological Order Parameters Far from Equilibrium, Phys. Rev. B {\bf93}, 085416 (2016).
    
    
    
    
    \bibitem{Uhlmann2007} Michael Uhlmann, Ralf Schützhold, and Uwe R. Fischer, Vortex Quantum Creation and Winding Number Scaling in a Quenched Spinor Bose Gas, Phys. Rev. Lett. {\bf99}, 120407(2007).
    \bibitem{Uhlmann2010} Michael Uhlmann, Ralf Schützhold, and Uwe R. Fischer, $O(N)$ symmetry-breaking quantum quench: Topological defects versus quasiparticles, Phys. Rev. D {\bf81}, 025017 (2010).
    \bibitem{Uhlmann2010b} Michael Uhlmann, Ralf Schützhold and Uwe R Fischer, System size scaling of topological defect creation in a second-order dynamical quantum phase transition, New J. Phys. {\bf12} 095020 (2010).
    \bibitem{Schützhold2006} Ralf Schützhold, Michael Uhlmann, Yan Xu, and Uwe R. Fischer, Sweeping from the Superfluid to the Mott Phase in the Bose-Hubbard Model, Phys. Rev. Lett. {\bf97}, 200601 (2006).
    
    
    
    
   \bibitem{Damski2005} B. Damski, The Simplest Quantum Model Supporting the Kibble-Zurek Mechanism of Topological Defect Production: Landau-Zener Transitions from a New Perspective, Phys. Rev. Lett. {\bf95}, 035701 (2005).
   
   
   
    
    
    \bibitem{Botet1983} R. Botet and R. Jullien, Large-Size Critical Behavior of Infinitely Coordinated Systems, Phys. Rev. B {\bf28}, 3955 (1983).
    
    
    
    
    \bibitem{Sharma2016} S. Sharma, U. Divakaran, A. Polkovnikov, and A. Dutta, Slow Quenches in a Quantum Ising Chain: Dynamical Phase Transitions and Topology, Phys. Rev. B {\bf93}, 144306 (2016).
    \bibitem{Puškarov2016} T. Puškarov and D. Schuricht, Time Evolution during and after Finite-Time Quantum Quenches in the Transverse-Field Ising Chain, SciPost Phys. {\bf1}(1), 003 (2016).
    \bibitem{Zamani} Sara Zamani, J. Naji, R. Jafari, A. Langari, Scaling and Universality at Ramped Quench Dynamical Quantum Phase Transition, arXiv:2310.15101.
    \bibitem{Cao} Kaiyuan Cao, Peiqing Tong, Exploring Dynamical Phase Transitions in the XY Chain through Linear Quench: Early and Long-term Perspectives, arXiv:2405.08449.
    \bibitem{Baghran} R. Baghran, R. Jafari, and A. Langari, Competition of long-range interactions and noise at a ramped quench dynamical quantum phase transition: The case of the long-range pairing Kitaev chain, Phys. Rev. B {\bf110}, 064302 (2024).
    
    
    
    \bibitem{Dusuel2005}S. Dusuel and J. Vidal, Continuous unitary transformations and finite-size scaling exponents in the Lipkin-Meshkov-Glick model, Phys. Rev. B {\bf71}, 224420 (2005).
    \bibitem{Acevedo2014} O. L. Acevedo, L. Quiroga, F. J. Rodríguez, and N. F. Johnson, New Dynamical Scaling Universality for Quantum Networks across Adiabatic Quantum Phase Transitions, Phys. Rev. Lett. {\bf112}, 030403 (2014).
    \bibitem{Titum2020} P. Titum and M. F. Maghrebi, Nonequilibrium Criticality in Quench Dynamics of Long-Range Spin Models, Phys. Rev. Lett. {\bf125}, 040602 (2020).
    
    
    
    \bibitem{Dabrowski2016} R. Dabrowski and G. V. Dunne, Time Dependence of Adiabatic Particle Number, Phys. Rev. D {\bf94}, 065005 (2016).
    \bibitem{Lewis1967} H. R. Lewis, Classical and Quantum Systems with Time-Dependent Harmonic-Oscillator-Type Hamiltonians, Phys. Rev. Lett. {\bf18}, 636 (1967).
    \bibitem{Defenu2018} N. Defenu, T. Enss, M. Kastner, and G. Morigi, Dynamical Critical Scaling of Long-Range Interacting Quantum Magnets, Phys. Rev. Lett. {\bf121}, 240403 (2018).
    
    
    \bibitem{sm} See Supplemental Materials for detailed discussions on the dynamical quantum phase transition in different models and on the argument of impulse-adiabatic scenario. 
    
    
    \bibitem{Mukherjee2010} V. Mukherjee and A. Dutta, Adiabatic Multicritical Quantum Quenches: Continuously Varying Exponents Depending on the Direction of Quenching, Epl {\bf92}, (2010).
    \bibitem{Deng2009} S. Deng, G. Ortiz, and L. Viola, Anomalous Nonergodic Scaling in Adiabatic Multicritical Quantum Quenches, Phys. Rev. B {\bf80}, 241109(R) (2009).
    \bibitem{Divakaran2008} U. Divakaran, A. Dutta, and D. Sen, Quenching along a Gapless Line: A Different Exponent for Defect Density, Phys. Rev. B {\bf78}, 144301 (2008).
    \bibitem{Porta2020} Porta, S., Cavaliere, F., Sassetti, M. et al. Topological classification of dynamical quantum phase transitions in the xy chain, Sci Rep {\bf10}, 12766 (2020).
    
    
    
    
    
    \bibitem{Zhou2018} L. Zhou, Q. H. Wang, H. Wang, and J. Gong, Dynamical Quantum Phase Transitions in Non-Hermitian Lattices, Phys. Rev. A {\bf98}, 022129 (2018).
    \bibitem{Jing2024} Y. Jing, J.-J. Dong, Y.-Y. Zhang, and Z.-X. Hu, Biorthogonal Dynamical Quantum Phase Transitions in Non-Hermitian Systems, Phys. Rev. Lett. {\bf132}, 220402 (2024).
    \bibitem{Shen2019} X. Shen, F. Wang, Z. Li, and Z. Wu, Landau- Zener-Stückelberg interferometry in PT-symmetric non- Hermitian models, Phys. Rev. A {\bf100}, 062514 (2019).
    \bibitem{Sim2023} K. Sim, N. Defenu, P. Molignini, and R. Chitra, Quantum Metric Unveils Defect Freezing in Non-Hermitian Systems, Phys. Rev. Lett. {\bf131}, 156501 (2023).
    \bibitem{Lee2015} T. Lee, U. Alvarez-Rodriguez, X. Cheng, L. Lamata, and E. Solano, Tachyon physics with trapped ions, Phys. Rev. A {\bf92}, 032129 (2015).
    \bibitem{Gerritsma2010} R. Gerritsma, G. Kirchmair, F. Zähringer, E. Solano, R. Blatt, and C. F. Roos, Quantum simulation of the dirac equation, Nature (London) {\bf463}, 68 (2010).
    
    
    
    
    
    
\bibitem{Fauseweh2024} Fauseweh, B. Quantum many-body simulations on digital quantum computers: State-of-the-art and future challenges. Nat Commun {\bf15}, 2123 (2024).
\bibitem{King2022} King, A.D., Suzuki, S., Raymond, J. et al. Coherent quantum annealing in a programmable 2,000 qubit Ising chain. Nat. Phys. {\bf18}, 1324–1328 (2022).

\bibitem{Li2023} B. W. Li et al., Probing Critical Behavior of Long-Range Transverse-Field Ising Model through Quantum Kibble-Zurek Mechanism, PRX Quantum {\bf4}, 010302 (2023).
\bibitem{Keesling2019} Keesling, A., Omran, A., Levine, H. et al. Quantum Kibble–Zurek mechanism and critical dynamics on a programmable Rydberg simulator. Nature {\bf568}, 207–211 (2019).
\bibitem{Xu} K. Xu et al., Probing Dynamical Phase Transitions with a Superconducting Quantum Simulator, Sci. Adv. {\bf6}, (2020).


\end{thebibliography}
 \end{document}